\documentclass[twocolumn,showpacs,preprintnumbers,amsmath,amssymb]{revtex4}

\usepackage{graphicx}% Include figure files
\usepackage{bm}% bold math

\begin{document}

\title{Response to targeted perturbations  for random walks on networks}

\date{\today}

\author{Vincent Tejedor}
\affiliation{Laboratoire de Physique Th{\'e}orique de la Mati\`ere
Condens\'ee, Universit{\'e} Pierre et Marie Curie, Paris, France}
\author{Olivier B{\'e}nichou}
\affiliation{Laboratoire de Physique Th{\'e}orique de la Mati\`ere
Condens\'ee, Universit{\'e} Pierre et Marie Curie, Paris, France}
\author{Raphael Voituriez}
\author{Michel Moreau}
\affiliation{Laboratoire de Physique Th{\'e}orique de la Mati\`ere
Condens\'ee, Universit{\'e} Pierre et Marie Curie, Paris, France}
\begin{abstract}

We introduce  a general framework, applicable to a broad class of random walks on networks,  that quantifies  the response of the mean first-passage time  to a  target node to a local perturbation of the network, both in the context of attacks (damaged link) or strategies of transport enhancement (added link). This approach enables to determine explicitly  the dependence of this response on geometric parameters (such as the network size and the localization of the perturbation) and on the intensity of the perturbation. In particular, it is showed that the relative variation of the MFPT is independent of the network size, and remains significant  in the large size limit. Furthermore,  in the case of non compact exploration of the network, it is found that  a targeted perturbation keeps a substantial impact on transport properties for any localization of the damaged link. 

\end{abstract}

%\pacs{?}

\maketitle

Since the pioneering works of Erd\"os and R\'enyi on random graphs \cite{erdos59}, complex networks have been extensively studied, either for their theoretical interest, or for their applications in various areas such as social and economical sciences, biology or  epidemiology (see for instance \cite{Albert:2002,Barratbook,Dorogovtsev:2008} and references therein). 
Among other features,  transport properties on networks play a crucial role, and in this context, random walks on networks have appeared as a prototypical example of dynamical process which has been extensively studied over the past few years  \cite{Noh:2004a,Bollt:2005,Baronchelli:2008,nechaev2002}. In particular, the target search problem has raised much attention because of the variety of its applications \cite{package_animaux}, and the mean first-passage time (MFPT) to a target node is the most commonly used quantitative indicator of the efficiency of a search process on a network \cite{Gallos:2007a,package_nature,Haynes:2008,Agliari:2009,Zhang:2009}. A general framework to calculate MFPTs on networks has been presented in \cite{Noh:2004a}  and since then explicit results have been obtained both for fixed \cite{package_nature} and averaged positions \cite{Haynes:2008,Agliari:2009,Zhang:2009,Tejedor:2009vn} of the starting point.

In parallel, increasing interest in modeling failures or attacks in networks has developed \cite{Moreira:2009ly}, for instance in the context of epidemic spreading \cite{Pastor01}, virus attack on networks \cite{Barabasi02}, or electrical blackout \cite{Dobson:2002,Buldyrev:2010ys}. Most of these studies deal with static properties of networks after a given perturbation (typically the  removal of nodes or links), with the notable exception of \cite{Simonsen:2008}. In particular it has been showed that scale free networks are very resilient to random perturbations, while the  targeted removal of a hub can have dramatic consequences.  Very recently, it has also been put forward that in the case of  interdependent networks a broader degree distribution increases the vulnerability to random failure \cite{Buldyrev:2010ys}.

In this context, quantifying the impact of targeted perturbations of a network on the search efficiency and more generally on transport properties is an important and yet unexplored problem. In this letter, we provide  a general framework that quantifies  the response of the MFPT to a  target node to a local perturbation of the network, both in the context of attacks (damaged link) or strategies of transport enhancement (added link). This approach enables to determine explicitly  the dependence of this response on geometric parameters (such as the network size and the position of the perturbation) and on the intensity of the perturbation. In particular, it reveals that  the relative variation of the MFPT is  independent of the network size $N$ in the large $N$ limit,  and remains significant  even for very large networks ; additionally,  in the  case of non compact exploration of the network, a targeted perturbation keeps a substantial impact on transport properties for any localization of the damaged link.

We consider a discrete time random walker on a network of $N$ nodes,  and assume that the transition 
probabilities $R_{wu}$ from $u$ to $w$ of 
 the walker are such that an stationary distribution $P_x$ exists. In addition, we denote $P^n_{xy}$ the probability to be at site $x$ after $n$ steps for a random walk starting from site $y$.
In what follows, we will be interested in the influence of  targeted perturbations of the network on  the MFPT  $\langle T_{TS} \rangle$ from a starting site $S$ to a target site $T$.

We first introduce the general method to study  the influence of  a single link perturbation on the MFPT, and take  the example of the weakening or removal of a link. More precisely,  we assume that the  transition probability $R_{wu}$ from $u$ to $w$ is changed by $\delta R_{wu}<0$. Without loss of generality, we here assume that this perturbation is compensated on the probability $R_{uu}$ to stay at $u$ during the elementary step, ie $\delta R_{uu} = - \delta R_{wu} $ , and that all other transition rates are  unchanged. Note that the important particular case of a broken link is then  given by taking   $\delta R_{wu}=- R_{wu}$. In the sequel, all quantities denoted with a prime correspond to the perturbed situation.

The MFPT  in the perturbed situation  can be calculated by first noting 
 that the perturbation affects only the trajectories that pass through $u$ before reaching $T$, so that, for any starting point $x$ :
 \begin{equation}
\label{mod}
\langle T_{Tx}' \rangle - \langle T_{Tx} \rangle = P_{ux}^T \left ( \langle T_{Tu}' \rangle - \langle T_{Tu} \rangle \right ),
\end{equation}
where $P_{ux}^T$ is the splitting probability to reach $u$ before $T$ starting from $x$. Writing next the backward equation for the MFPT \cite{Redner:2001a} :
\begin{equation}
\langle T_{Tu}' \rangle = 1 + \sum_{v} R_{vu}' \langle T_{Tv}' \rangle
\end{equation}
both in the perturbed and unperturbed situations, we obtain
\begin{eqnarray}
\label{2}
\langle T_{Tu}' \rangle - \langle T_{Tu} \rangle & = & \sum_{v} R_{vu} \left ( \langle T_{Tv}' \rangle - \langle T_{Tv} \rangle \right ) \nonumber \\ 
& & + \delta R_{wu} (\langle T_{Tw}' \rangle - \langle T_{Tu}' \rangle).
\end{eqnarray}
Using Eqs(\ref{mod}),(\ref{2}), the variation of the MFPT starting from site $u$ is found to be given by :
\begin{equation}
\langle T_{Tu}' \rangle - \langle T_{Tu} \rangle = \frac{\delta R_{wu}(\langle T_{Tw} \rangle - \langle T_{Tu} \rangle)}{1-\bar{P}_{uu}^T+\delta R_{wu}(1-P_{uw}^T)} ,
\end{equation}
where we have introduced $\bar{P}_{uu}^T \equiv \sum_v R_{vu} P_{uv}^T$, defined as  the probability to come back to  $u$ before reaching $T$. 
Using (\ref{mod}), we  finally obtain the relative MFPT variation $\delta_{TS}\equiv (\langle T_{TS}' \rangle-\langle T_{TS} \rangle)/\langle T_{TS} \rangle $ for any starting site $S$:
\begin{equation}
\label{lienexact}
\delta_{TS} =  \frac{P_{uS}^T}{\langle T_{TS} \rangle}\frac{\delta R_{wu}(\langle T_{Tw} \rangle - \langle T_{Tu} \rangle)}{1-\bar{P}_{uu}^T+\delta R_{wu}(1-P_{uw}^T)} .
\end{equation}
Using \cite{package_nature, Sylvain_package}, $\delta_{TS}$ can be expressed as a function of  the perturbation $\delta R_{wu}$, the unperturbed stationary distribution $P_x$,  and  the pseudo-Green function $\displaystyle H_{xy}\equiv \sum_{n=0}^\infty (P^n_{xy}-P_x)$ of the unperturbed problem, giving :
\begin{eqnarray}
\label{lienexact2}
&&\!\!\!\!\!\delta_{TS}=\frac{H_{Tu}-H_{Tw}}{H_{TT}-H_{TS}}\times\nonumber\\
&&\!\!\!\!\!\!\!\!\!\!\!\!\!\!\!\times\frac{\delta R_{wu}[P_T(H_{uS}-H_{uT})+P_u(H_{TT}-H_{TS})]}
{P_T+\delta R_{wu}[P_T(H_{uu}-H_{uw})+P_u(H_{Tw}-H_{Tu})]}.
\end{eqnarray}
This central result has several important implications. 

First,  we stress that in  the particular case of a regular $d$--dimensional hypercubic  parallelepipedic network with constant probability transitions between nearest neighbors  the pseudo Green functions are known exactly \cite{Barton:1989}.  Using next that the stationary probability is in this case uniform ($P_x=1/N$ for any node $x$), Eq.(\ref{lienexact2})  provides  an exact and fully explicit result for the effect of an arbitrary modification of a given link on the MFPT.

\begin{figure}[h!]
\includegraphics[width =\linewidth,clip]{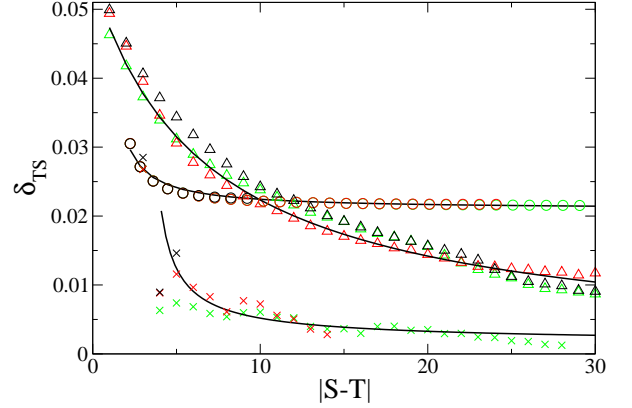} 
\caption{Relative variation of the MFPT as a function of the distance $|S-T|$ between source and target (here the relative positions of $T$, $u$ and $w$ are fixed and averages over triplets are taken). Numerical simulations (symbols) are compared to the approximation (\ref{lienexact2},\ref{scaling}) (plain lines) for 3 different network sizes.  Circles stand for 3D regular lattice ($d_w-d_f = -1$), triangles for 2D critical percolation cluster ($d_w-d_f \simeq 0.98$), crosses for a $(2,2)$ flower ($d_w = d_f$).}
\label{fig:MFPT}
\end{figure}
\begin{figure}[htb!]
\includegraphics[width =\linewidth,clip]{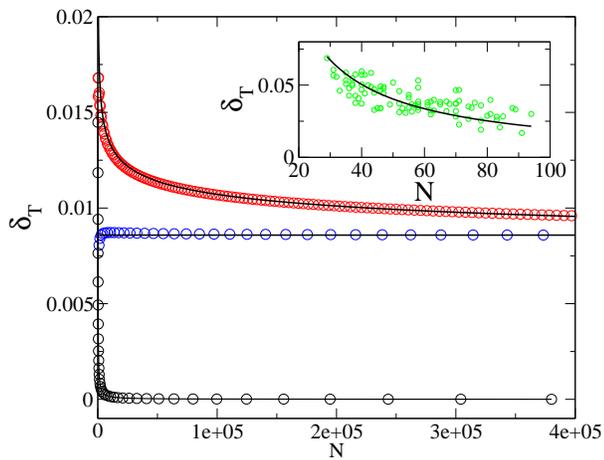} 
\caption{Relative variation of the MFPT averaged over $S$, $\delta_T$ for different network sizes $N$ for 1D (black), 2D (red) and 3D (blue) cubic lattices and a 2D critical percolation cluster (green). The relative positions of $T$, $u$ and $w$ are fixed for all networks of a given kind. The circles stand for the simulated $\delta_T$, the black lines for the theoretical prediction (\ref{scaling2}),  where $C$ is a fitting parameter. }
\label{fig:GMFPT}
\end{figure}

Second,  in the  case of more general networks, but possessing  scale-invariant properties,  the dependence on the geometrical parameters can still be determined, by taking the large network size limit.  More precisely, this limit can be conveniently discussed when the random walk is a scale-invariant process, ie when the infinite volume propagator satisfies $P^n_{xy}\propto n^{-d_f/d_w}\Pi(|x-y|/n^{1/d_w})  $ where  $d_w$ and $d_f$ denote respectively the walk dimension and the fractal dimension of the network, and where $|x-y|$ denote the distance between nodes $x$ and $y$. Indeed, in this case all the differences $H_{xy}-H_{xz}$ that enters Eq. (\ref{lienexact}) can be rewritten in terms of differences $H_{xx}-H_{xy}$ which satisfy in the large volume limit \cite{package_nature}:
\begin{equation}\label{scaling}
H_{xx}-H_{xy} \sim \left \{ \begin{array}{ll}
A+  B|x-y|^{d_w-d_f}  &  \textnormal{ if } d_w < d_f\\
A+ B \ln |x-y|  &  \textnormal{ if } d_w = d_f\\
B|x-y|^{d_w-d_f}  &  \textnormal{ if } d_w > d_f
\end{array} \right .
\end{equation}
where $A,B$ are numerical constants depending only on the  infinite volume propagator. 
Equations (\ref{lienexact2})-(\ref{scaling}) have two consequences. (i) The independence of $N$ of Eq.(\ref{scaling}) readily gives that the relative variation   $\delta_{TS}$ is {\it independent of} $N$ in the large volume limit. This quite unexpected effect,  illustrated by the data collapse for different volumes in Fig. \ref{fig:MFPT} on various examples of networks, implies that the effect of a targeted perturbation is not diluted but remains finite even for extremely large networks.  Actually this universal asymptotic independence on $N$ of the relative variation  $\delta_{TS}$ for fixed starting node $S$ strongly differs from the relative variation  $\delta_{T}$ of the MFPT averaged over $S$ defined by $\langle T_T\rangle\equiv\sum_S P_S \langle T_{TS}\rangle$.  More precisely, assuming next detailed balance, it can be shown using symmetry relations of $H_{xy}$  that
\begin{eqnarray}
&&\delta_{T}\equiv\frac{\langle T_{T}' \rangle-\langle T_{T} \rangle}{\langle T_{T} \rangle}
=\frac{H_{Tu}-H_{Tw}}{H_{TT}}\times\nonumber\\
&&\!\!\!\!\!\!\!\!\!\!\!\!\!\!\!\times\frac{\delta R_{wu}P_u(H_{TT}-H_{Tu})}
{P_T+\delta R_{wu}[P_T(H_{uu}-H_{uw})+P_u(H_{Tw}-H_{Tu})]}.
\end{eqnarray}
In the large volume limit, this exact expression leads to 
\begin{equation}\label{scaling2}
\delta_{T} \sim \left \{ \begin{array}{ll}
C  &  \textnormal{ if } d_w < d_f\\
C/ \ln N  &  \textnormal{ if } d_w = d_f\\
C/N^{d_w/d_f-1}  &  \textnormal{ if } d_w > d_f
\end{array} \right.
\end{equation}
where we have derived the asymptotic expression of $H_{TT}$ using \cite{Tejedor:2009vn,BenichouO:2010a}.
In other words, as for the relative variation $\delta_{T}$, the $N$ independence is recovered only in the  case  $d_f>d_w$ of so-called non compact exploration, while a strong dependence on $N$ is found in the opposite case of compact exploration ($d_w\ge d_f$).  This effect is illustrated in Fig. \ref{fig:GMFPT}.
 (ii) The asymptotic form (\ref{scaling}) used in Eq.(\ref{lienexact2})  also provides  the explicit dependence of $\delta_{TS}$ on the relative distances between the nodes  $S,T,u,w$.
Such dependence has been checked numerically  (see  Fig.  \ref{fig:MFPT}) for various networks such as regular euclidian lattices ($d_w = 2$), 2D critical percolation clusters ($d_w = 2.88$ and $d_f = 91/48$), and $(u,v)$-flowers which  are recursive fractals defined in \cite{Rozenfeld:2007}.  Fig.  \ref{fig:MFPT} reveals a very  different behavior in the case of compact exploration (illustrated by  critical percolation clusters)  for which $\delta_{TS}$ vanishes at larges distances,  and in the case of  non-compact exploration (illustrated by 3D regular lattices), for which $\delta_{TS}$ remains finite even for very large distances. Noteworthily this shows that in the non compact case a targeted perturbation keeps a substantial impact on transport properties for {\it  any localization} of the damaged link. 

\begin{figure}[h!]
\includegraphics[width =\linewidth,clip]{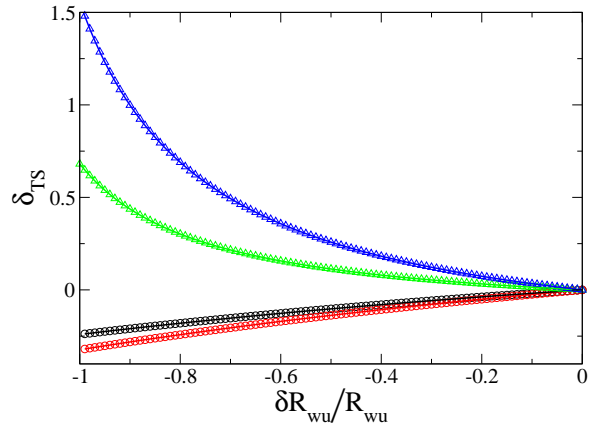}
\caption{Relative variation of the MFPT as a function of the $u \to w$ link perturbation, for given $T$, $u$ and $w$ sites, and two different $S$ for each network : 2D critical percolation clusters (circles) and  $(3,3)$-flower of generation 3 (triangles). For the flower network, the perturbed link leads to $T$. Numerical simulations are fitted with  Eq. (\ref{eq:fit1pt}).}
\label{fig:1pt}
\end{figure}

%More surprising, the rough approximation of equation (\ref{lienapprox}) is very good: the occupation time is often quite small, and gets smaller if the network gets bigger. 
%We tried this formula in varied networks : a 2D quadratic network, a $(2,2)$- and a $(3,3)$-flowers. Those flowers are recursive fractals obtain with the following procedure : at each step, every link is cuted, and replaced by two path of $2$ (resp. $3$) links (with a node between two new links). Since approximations of the Green function are known, we have a $0$ constant formula. In average, the formula (\ref{lienapprox}) works very well ({\it cf.} figure \ref{caslimite} and \ref{caslimite2}) ! 

Third, in the general case of a random walk on an arbitrary network,  where the pseudo-Green functions can be difficult to evaluate, Eq. (\ref{lienexact2}) still gives explicitly the functional  dependence of $\delta_{TS}$ on the perturbation $\delta R_{wu}$, which takes the form
\begin{equation}
\delta_{TS} =  \frac{D \delta R_{wu}}{E+\delta R_{wu}}
\label{eq:fit1pt}
\end{equation}
where $D$ and $E$ do not depend on $\delta R_{wu}$ (note that $E$ does not depend on $S$ either). This general form has been validated by numerical simulations in   Fig. \ref{fig:1pt}. Additionally, provided that the differences $H_{xx}-H_{xy}$ involved in Eq. (\ref{lienexact2}) have a finite limit in the large volume regime, $D$ and $E$ turn out to be independent of $N$. This shows that the independence of  $\delta_{TS}$ on $N$ still holds in this case, and makes this property very robust.

Finally,  it should be noted that the relative variation $\delta_{TS}$ remains rather weak for an arbitrary perturbed link. We however stress that such local attack of a network is not affected by dilution effects and  remains finite even in the large volume limit;  additionally, in the non compact case, it is also widely independent of the localization of the perturbation and non vanishing even for a very remote damaged link. Furthermore, the effect of a local perturbation can  become much stronger if targeted  to a link directly leading to the target, %(encompassed by  Eq. (\ref{lienexact2})),  
as can be expected intuitively (see  Fig. \ref{fig:1pt}).

\begin{figure}[htb!]
\includegraphics[width =\linewidth,clip]{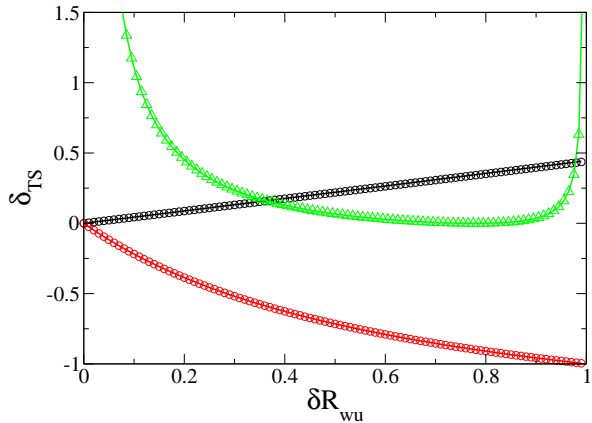} 
\caption{Relative variation of the MFPT in response to the addition of a new link, in the directed (red and black lines and symbols) and bidirectional case (green line and symbols) for  3D regular lattices. Simulations (symbols) are plotted against the theoretical prediction (Eq. (\ref{newlink}) and its extension, plain lines).  Black circles: $S$, $T$, $u$ and $w$ are distinct, and the new link starts from a target neighbor. Red circles:  the new link is between $S$ and $T$. Green triangles : addition of a link 
 between two 3D regular lattices (with $\delta R_{wu} = \delta R_{uw}$), $S$ being in the first lattice, and $T$ in the
second. Here we  define $\delta_{\textnormal{ST}} =
(\textnormal{MFPT}-\min(\textnormal{MFPT}))/\min(\textnormal{MFPT})$. }
\label{creationt}
\end{figure}

Importantly,  the above formalism can be extended  to tackle the reciprocal problem of enhancing instead of damaging the transport abilities of a network. As a first step in this direction, we quantify the effect  of adding a new link between two nodes. The definition of the perturbation has to be slightly modified in the case of an added link, since the case studied above is ill defined for $\delta R_{wu}>0$ and $R_{uu}=0$.  We assume now that there is initially no link  between $u$ and $w$ ($R_{wu}=0)$, and consider a perturbation $\delta R_{wu}>0$. In turn,  we set for all neighbors $v$ of $u$ that 
$\delta R_{vu}= -\delta R_{wu}/k(u)$, where $k(u)$ is the initial connectivity of $u$ (note that  $\delta R_{wu}$ is assumed to be small enough so that all transition probabilities are positive). In this case, the equivalent of Eq. (\ref{lienexact}) can be  obtained along the same line and reads:
\begin{equation}
\delta_{TS}  =   \frac{P_{uS}^T}{\langle T_{TS} \rangle} \frac{\delta R_{wu} \left ( \langle T_{Tw} \rangle - \langle T_{Tu} \rangle + 1 \right )}{1-\bar{P}_{uu}^T + \delta R_{wu} \left ( \bar{P}_{uu}^T - P_{uw}^T \right )} , 
\label{newlink}
\end{equation}
which, as previously,  can be expressed only  in terms of $\delta R_{wu}$  and pseudo-Green functions. This theoretical prediction is plotted against  numerical simulations on the example of   parallelepipedic networks in Fig. \ref{creationt} (black and red lines).  Note that the MFPT can be decreased very significantly  if the new link points to the target (red line of Fig. \ref{creationt}). 
 Last, we stress that  the case of  a single added bidirectional  link connecting two nodes $u$ and $w$ of initially distinct networks ${\cal N}_1, \  {\cal N}_2$ can be obtained using the same method, and yields an explicit results for the MFPT which is displayed in Fig. \ref{creationt} (green line). This constitutes a first step in designing interdependent networks as introduced in \cite{Buldyrev:2010ys}.

To conclude, we have presented  a general framework that quantifies  the response of the MFPT to a  target node to a local perturbation of the network, both in the context of attacks (damaged link) or strategies of transport enhancement (added link). This approach enables to determine explicitly  the dependence of this response on geometric parameters (such as the network size and the position of the perturbation) and on the intensity of the perturbation. It reveals that the relative variation of the MFPT is independent of the network size, and remains significant  even in the large size limit. Additionally,  in the non compact case a targeted perturbation keeps a substantial impact on transport properties for any localization of the damaged link.

%\bibliography{biblio}

\end{document}